\documentclass[iicol,a4]{sn-jnl}% Math and Physical Sciences Reference Style
\usepackage{tabularx}
\usepackage{xspace}
\usepackage{subcaption}
\usepackage[inline]{enumitem}

\newcommand{\ag}{Ag-valuate\xspace}
\newcommand{\ak}{AgAsk\xspace}
\newcommand{\ra}[1]{\renewcommand{\arraystretch}{#1}}		% booktabs
\jyear{2022}%

\theoremstyle{thmstyleone}%
\theoremstyle{thmstyletwo}%

\theoremstyle{thmstylethree}%

\begin{document}

\title{AgAsk: An Agent to Help Answer Farmer's Questions From Scientific Documents}

\author*[2]{\fnm{Bevan} \sur{Koopman}}\email{bevan.koopman@csiro.au}
\equalcont{These authors contributed equally to this work.}

\author*[1]{\fnm{Ahmed} \sur{Mourad}}\email{a.mourad@uq.edu.au}
\equalcont{These authors contributed equally to this work.}

\author[1]{\fnm{Hang} \sur{Li}}\email{hang.li@uq.edu.au}
\author[1]{\fnm{Anton van der} \sur{Vegt}}\email{a.vandervegt@uq.edu.au}
\author[1]{\fnm{Shengyao} \sur{Zhuang}}\email{s.zhuang@uq.edu.au}
\author[2]{\fnm{Simon} \sur{Gibson}}\email{simon.gibson@csiro.au}
\author[1]{\fnm{Yash} \sur{Dang}}\email{y.dang@uq.edu.au}
\author[3]{\fnm{David} \sur{Lawrence}}\email{david.lawrence@daf.qld.gov.au}
\author[1]{\fnm{Guido} \sur{Zuccon}}\email{g.zuccon@uq.edu.au}

\affil*[1]{\orgname{The University of Queensland}, \city{Brisbane}, \country{Australia}}

\affil[2]{\orgname{CSIRO}, \city{Brisbane}, \country{Australia}}

\affil[3]{\orgname{Queensland Department of Agriculture and Fisheries}, \city{Brisbane}, \country{Australia}}

\abstract{
	Decisions in agriculture are increasingly data-driven; however, valuable agricultural knowledge is often locked away in free-text reports, manuals and journal articles. 
Specialised search systems are needed that can mine agricultural information to provide relevant answers to users' questions. 
This paper presents AgAsk --- an agent able to answer natural language agriculture questions by mining scientific documents.

We carefully survey and analyse farmers' information needs. On the basis of these needs we release an information retrieval test collection comprising real questions, a large collection of scientific documents split in passages, and ground truth relevance assessments indicating which passages are relevant to each question.

We implement and evaluate a number of information retrieval models to answer farmers questions, including two state-of-the-art neural ranking models. We show that neural rankers are highly effective at matching passages to questions in this context.

Finally, we propose a deployment architecture for AgAsk that includes a client based on the Telegram messaging platform and retrieval model deployed on commodity hardware.

The test collection we provide is intended to stimulate more research in methods to match natural language to answers in scientific documents. While the retrieval models were evaluated in the agriculture domain, they are generalisable and of interest to others working on similar problems.

The test collection is available at: \url{https://github.com/ielab/agvaluate}.
}

\keywords{Information Retrieval, Professional Search, Domain-specific Search, Agriculture, Passage Retrieval}



\maketitle

\section{Introduction}
\label{sec:intro}

Twenty first century agriculture is increasingly mechanised, data-driven and scientific-evidence based~\cite{smith2018getting,virgona2011evidence,bacco2019digitisation}. Even developing countries are seeing increasing digital disruption~\cite{Jain:2018aa,Opoku-Agyemang:2017aa}.

% Describe the problem
A wealth of valuable resources and data %--- from research studies, project reports and communications, through to meteorological and soil sample data --- 
could be used by agricultural users, but there are significant barriers in  effectively accessing these resources. 

Much of these resources are currently locked away into large and heterogeneous datasets, research project reports, communications and scientific publications, meteorological and soil sample data, and external services and applications~\cite{lokers2016analysis}. Some are structured data; while large amounts are still in natural language form. 
These natural language documents are not easily discoverable and synthesised. No federated service is in place that offers agricultural users a single entry-point to search this type of information.
Thus agricultural users are not able to put into practice valuable insights from such information.

% IR specific problem
On the other hand, digital connectivity is not the major barrier to accessing agricultural resources. Farmers now make use of handheld devices and digital services, Twitter being one popular platform for farmers to keep informed of the latest trends~\cite{mills2019use,van2011iphone}. 

The real barrier is how to effectively serve farmers complex, multi-faceted information needs. Scientific-like questions such as ``What varieties of bread wheat are most resistant to crown rot?'' are hard to answer automatically. Two problems make these questions hard to answer:
\begin{itemize}
	\item \textbf{Complex answer matching}: Growers may express their queries in ways that do not directly match relevant information. The complex information need also comes with many variations in how users would express such need (query variations) and an automated system must handle such variations in a robust manner.
	\item \textbf{Focused answers}: Growers need easily digestible answers to their questions; presenting a 25 page scientific document will not do, both from a workload perspective and for grower's to recognised how it might relate to their query.
\end{itemize}

% The language in grower's queries may differ considerably from that of relevant information they wish to find. Growers may describe, for example, a pest via visible symptoms, with the aim of first identification and finally of deciding upon a course of action (treatment). The issue here is how to automatically infer the semantic association between the language of the query and the language of the relevant information.  

% The level of technical expertise of the growers varies too: some growers may pose queries using specific technical terms that match those of scientific sources they are seeking (e.g., use of Latin species names), while others will use lay or even colloquial terms. 
%A conversational agent needs to handle both.

% Current IR solutions
The above are common IR problems for which there are some existing solutions. For complex answer matching, neural models are currently state-of-the-art~\cite{Craswell:2022aa}. These methods do not rely on matching individual terms but instead rely on learned representations of word meaning. This breaks the dependence on specific terms used in queries and relevant passages and allows for `semantic' matching, that is matching based on word meaning.

For producing focused answers, breaking documents into passages and ranking these against a user's query can provide the digestible answers users seek. Again, neural methods encode short passages of text into a representations that can be effectively ranked against a user's query --- another short passage of text itself.

% The Gap

% How we will fix it

\subsection*{Contributions}

This paper presents a framework that serves the information needs of agricultural users. It:
\begin{itemize}
	\item Analyses the information needs of real agricultural users, including the sources of information they use.
	
%	\item Surveys the literature for information technology solutions used in agriculture that involve some form of search;
%	\item Conducts a user survey of real agriculture users to understand both their information needs and where they currently look for answers;
	\item Provides a test collection in a new and growing domain --- search from scientific articles in general and in the agriculture domain in particular --- which comprises a 86,846 document collection (further divided into 9,441,693 passages) carefully compiled by domain experts rather than web crawling or crowd sourcing. It also provides 210 rich, multi-faceted, real-world search topics comprising: i) a natural language question; ii) multiple keyword query variations; iii) an expert-authored answer; and iv) graded relevance assessment of passages.
	\item Provides a series of retrieval experiments with both baseline term-based retrieval models and state-of-the-art neural rankers. %Some insights into how models pre-trained in a general domain can be fine-tuned to a special domain are also provided.
	\item Provides an end to end system, \ak, that offers agricultural users a single entry-point to search this information.
%	\item A brief explanation of a number of different tasks and research directions that could be explored with \ak and its associate data to meet the information needs of agricultural users.
\end{itemize}

These contributions touch on all aspects of the problem: from the needs of the users to the resources required to investigate the problem, the underlying machine learning model and a production search system.

\section{Related Work}
 
While conversational agents have been proposed as a viable means to provide good answers to growers' questions~\cite{bacco2019digitisation,smith2018getting}, a limited number of solutions have been  proposed and explored.

A number of systems arose as a results of the release of a substantial dataset of farmer questions from the Kian Call Center (KCC).\footnote{\url{https://data.gov.in/dataset-group-name/kisan-call-centre}} KCC was a phone helpline service for farmers to consult with agriculture expert advisors about best practice and it was specifically tailored for the Indian market and agricultural context. Systems that used some portion of this dataset include AgriBot~\cite{jain2019agribot}, FarmChat~\cite{Jain:2018aa} and Krushi~\cite{momaya2021krushi}.

Agribot was developed to address growers information needs related to weather, market rates, plant protection and government funding opportunities. This conversational agent focused on the data of all Indian states collected over a 5 years period, and relied on sentence embeddings (sent2vec~\cite{arora2017simple}) and entity extraction to compute the similarity between a user question and a background of common question-answer pairs. Answers were sourced from an underlying agricultural knowledge base. Thus, unlike AgAsk, the knowledge base was not backed by a comprehensive collection of scientific evidence while requiring the manual curation of a domain-specific knowledge-base.

FarmChat was a speech-based conversational system that relied on decision rules and answers manually derived from the KCC data to identify answers, and on the IBM Watson APIs to perform intent identification and dialogue flow management. Much of the attention in FarmChat was on information access in a context of limited literacy and technology expertise in rural Ranchi, India, and on the information delivery modality (audio vs audio+text). FarmChat focused only on one crop (potatoes); it did not leverage machine learning for extracting knowledge but instead relied on a manually built knowledge base. The drawback of this approach is that FarmChat did not scale easily and was difficult to maintain and link to information sources. While it helped to answer grower questions similar to AgAsk, it was highly tailored to one crop and one region; unlike AgAsk which is both crop and region agnostic.

Krushi was a conversational chatbot aimed to address growers information needs related to weather, plant protection, animal husbandry, market price, fertiliser use, government schemes and soil testing. This conversational agent focused on the data of the nine districts in Maharashtra, India collected over a year. It utilised the RASA X conversational AI system, involving intent identification followed by response retrieval. It was made accessible to farmers via WhatsApp. 

Besides Indian resources which facilitate access to agricultural data that support farmers in rural areas, other resources have been developed in other countries. A user study collected 1,000 Taiwanese conversations from interviews between investigators and farmers discussing specific topics and was developed to address sales, logistics and plants~\cite{Chen:2021wi}. This data was utilised to train a LSTM sequence-to-sequence conversational model which relied on word embeddings to generate an answer to the input question.

Another resource developed a crop protection information system to support farmers in rural areas of Tanzania where it is hard for government agricultural officers to visit in a timely manner during seasonal diseases outbreaks~\cite{tende2021proposal}. A collection of 2,100 Swahili queries were gathered from face-to-face interviews with 100 farmers. The authors analysed farmers' preferred method of expressing their information needs (keyword queries or natural language questions, and via SMS or the Web). They showed that there is a significant association between the age of farmers and their preferred method for expressing their information need, with the majority of young farmers ($<$ 40) preferring short and simple SMS queries while old farmers preferring natural language questions. 

While all the aforementioned resources help advance the digitisation of agriculture, they have few key limitations: ; 1) they are limited to either specific regions or crops, which is not generalisable; 2) the question-answer pairs are not grounded to the source of information (e.g., a research article), which means there will be no responses for unseen questions; 3) scalability is hampered by manual curation of the data rather than leveraging machine learning for extracting knowledge~\cite{jain2018farmchat,tende2021proposal}.

Despite the increasing availability of rich data resources for farmers to draw on, there is a dearth of search-based systems that can brings this data together to answer a farmers query. The few examples of such search-based agents in the agricultural sector, although limited in scope, showed promise and indicate that a larger effect in this area would be fruitful. 

AgAsk addresses many of the aforementioned limitations by 1) being crop, region and question type agnostic; 2) being backed by a large collection of rigorous scientific information; 3) using automated methods to extract information, making the system scalable and avoiding manual curation; and 4) using state-of-the-art neural ranking models to match users questions to relevant passages (not documents) in the collection.

\section{Information Needs of Users in Agriculture}
\label{sec:info_needs}

%Creating the test collection involved the following steps: 1) gathering information needs; 2) creation of the documents collection; 3) creation of a set of topics (i.e., questions/queries); 4) forming a pool of passages for assessment; and 5) the relevance assessment process.\footnote{Ethics was granted by the University of Queensland for application \#2020000826.}

Users in agriculture can be broadly categorised into three types: growers (farmers), agronomists and specialists. 
The latter two are the experts that provide support to the farmers (either through paid consultations or sponsored by the government) and communicate to them the outcomes of recent research.
Each of these will have different information needs. 

In this section, we first survey the literature on the information needs of growers. Then, we conduct an online survey to gather the information needs of all three types.
The learning and materials detailed in this section will feed into the creation of an agricultural-specific test collection designed to evaluate search systems in this context; this is presented in Section~\ref{sec:collection}.
%, we compile a set of search topics based on the identified information needs. These topics provide a resource to empirically evaluate a number of proposed search models presented later in the paper, as well as a public resource for others doing research in this area.

\subsection{Types of Information Needs}
\label{sec:lit_info_needs}

From the literature we summarise the specific information needs of growers. We constrain our analysis to those farmers involved in crop production (i.e., growers) and exclude animal production. While much of the concepts outlined here are relevant to both, animal production includes substantial veterinary content, excluded for the benefit of brevity.

%The Kisan Call Center (KCC) provides a large dataset to derive insights into the information needs of growers. Although this data is specific to farmers in India, many of the underlying information needs captured there would be generally applicable in other contexts. An analysis of this dataset showed the top 5 query types were for pest and disease (61\%), weather (14\%), best practices (7\%), fertiliser use (5\%), and seeds (4\%)~\cite{Jain:2018aa}. Other surveys of growers show similar rankings of query types: plant protection and disease, marketing, fertiliser and water management, preparation of seedlings and sowing, and harvesting technology~\cite{Chauhan2012Information-hun}.

From the literature, \cite{Chauhan2012Information-hun,smith2018getting,Jain:2018aa} some key categories of information needs were identified and are outlined below.

\subsubsection*{Crop protection}
A significant number of grower's questions relate to protecting their crop from diseases or pests, whether for future prevention or because of an existing outbreak. 
In the latter cases, farmers often describe crop diseases via visible symptoms (e.g., ``brown spots on the leaves'') in order to first identify the diseases and second determine the best course of treatment (e.g., what fungicide to use, including dosage and application instructions). 
Similarly they may describe pest species (e.g., ``2cm black and yellow snail'') to determine the relevant pesticide to use. 
Many queries relate to identifying and eradicating weeds~\cite{Chauhan2012Information-hun}.
For all these queries, it is important to point out that the grower's query typically does not contain keywords that match the relevant answer (e.g., the actual pest species name); instead this needs to be inferred from the description of the symptoms / problems.

\subsubsection*{Best practices}
Growers are constantly on the lookout for how they can increase the quantity or quality of their yield as well as reduce their costs or wastage, consequently increasing profitability. Agriculture is constantly evolving with new products and practices; many growers feel that keeping abreast of current best practice is critical~\cite{smith2018getting}. While growers will ask specific questions on a topic when they require information, they also seek out recommendation services that ``push'' relevant information. For example, the use of Twitter is one common way of keeping abreast of trends~\cite{mills2019use,van2011iphone}.

\subsubsection*{Unbiased Product Recommendations}
Growers rely heavily on many agriculture products to run their farms. These can constitute a significant expense and as such they would like reliable and trustworthy product recommendations. Recommendations for different types of fertiliser, seed and crop variants and herbicide or pesticide are some commonly sought examples~\cite{Jain:2018aa}. 

\subsubsection*{Markets and Weather/Climate}
While the market and weather are factors outside grower's control, they will certainly wish to understand and adapt their practices to changes in both. Because a farm is a business producing agricultural products, it has the same requirements of access to and understanding of markets that all businesses have. Growers would like to understand and adapt to the market in which they operate~\cite{Chauhan2012Information-hun}. This includes understanding of current and projected prices on products they sell as well as costs of products and services they consume. 

Growers would like to take into account the past, current and future weather and climate. Planting, for example, is often tied specifically to periods of rainfall. Similarly, pest outbreaks often relate to weather and climate. Thus growers would want any information returned to be tailored to the recent weather. Similarly, upcoming weather impacts grower's decisions so information should be tailored to weather forecasts. Longer term climate information --- both historic and projected --- is also important to growers and needs to inform what information is presented to them.

\subsection{Understanding what Growers Want}
\label{sec:online_info_needs}

\begin{figure}
	\centering
	\includegraphics[width=0.9\columnwidth]{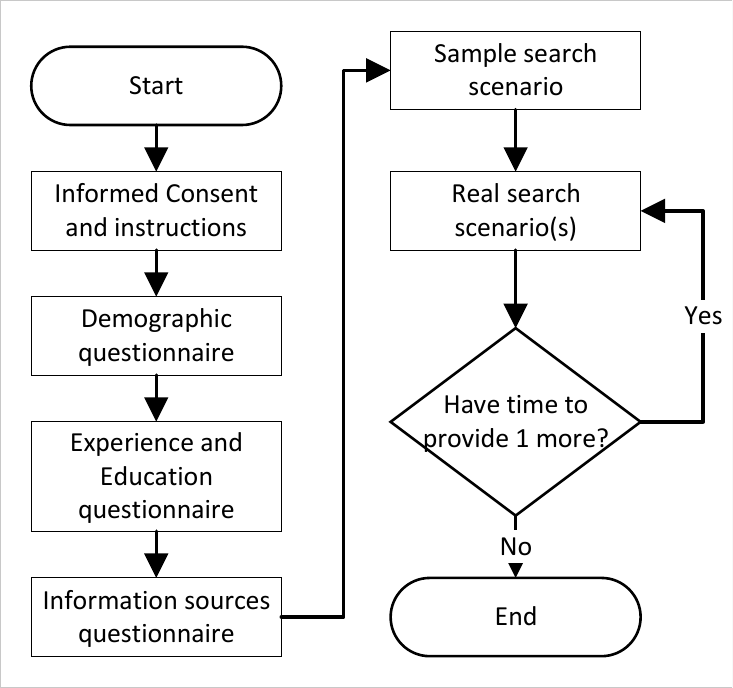}
	\caption{The online survey flowchart.}
	\label{fig:survey}
\end{figure}

\begin{table*}
	\centering
	\caption{The sample search scenario questionnaire.}
	\label{tbl:sample_search_scenario}
	\resizebox{\textwidth}{!}{
	\begin{tabular}{p{0.05\textwidth} p{0.95\textwidth}}
		\toprule
		\multicolumn{2}{l}{\textbf{Sample Question:} How much nitrogen fertiliser will I need to put on my crop this year, following the drought?}\\
		\midrule
		\multicolumn{2}{l}{In relation to the sample question:}\\
		\textbf{Q1.} & \ttfamily How important is answering the question to farm or crop success?\\
		\textbf{Q2.} & \ttfamily How frequently does it arise?\\
		\textbf{Q3.} & \ttfamily How urgent is it that you get an answer in a timely manner after it has arisen?\\
		\textbf{Q4.} & \ttfamily Select the top 5 sources of information you would go to in order to help you derive an answer to the sample question\\
		\textbf{Q5.} & \ttfamily Assuming that you had to search on the GRDC website or search on Google for an answer, please type in at least 3 different search queries that you would use to find information to help you.\\
		\textbf{Q6.} & \ttfamily Write down at least 3 elements of the answer that you would like to see as part of the complete answer to the question.\\
		\textbf{Q7.} & \ttfamily How much information would you like to receive in the answer\\
		\textbf{Q8.} & \ttfamily How might the answer be contextualised much more to your situation? List at least 3 additional specific information items\\
		\textbf{Q9.} & \ttfamily Write down a short summary answer, in 1-2 lines, if you know it.\\
		\textbf{Q10.} & \ttfamily If this is a question you have previously sought an answer to in real life, how successfully was the answer provided?\\
		\bottomrule
	\end{tabular}
	}
\end{table*}

To gain an accurate understanding of growers information needs and the resources they use to answer these,
%It was important that the information needs were derived from real users. Face-to-face interviews were not possible during the Covid pandemic; therefore, 
we conducted an online survey that targeted farmers, agronomists and agricultural specialists in Australia. %\footnote{Ethics was granted by the University of Queensland for application \#2020000826.} 
Figure~\ref{fig:survey} depicts the flowchart of the survey. The survey questionnaire solicited information about the users' demography, prior experience and education, information sources, and real search scenarios. The survey focused on grain crops based on the information needs from literature that growers care mostly about crop protection from diseases or pests. 
Participants were shown a sample search scenario related to grain growing questions that they might have had in the past 12 months and had to seek an answer to; i.e., questions that they could not answer with their own knowledge. Then they were asked to provide at least 3 and up to 5 real search scenarios. The sample search scenario presented to all participants is shown in Table~\ref{tbl:sample_search_scenario}.
While the number of participants is not representative, we share some of the insights that influenced our decisions for building the test collection, \ag.
%The questionnaire and results of this survey are made available online.\footnote{survey questionnaire and results}

\begin{table}
	\centering
	\caption{Statistics of the information needs survey.}
	\label{tbl:survey_stats}
	\resizebox{\columnwidth}{!}{
	\begin{tabular}{lr}
		\toprule
		\bf Role & 16 \\
		\bf \hspace{5pt} Grain grower & 0 \\
		\bf \hspace{5pt} Grain crop specialist & 9 \\
		\bf \hspace{5pt} Agronomist (farm consultant) & 7 \\
		\midrule
		\bf Years of experience &  16 \\
		\bf \hspace{5pt} 10 years or more         & 10 \\
		\bf \hspace{5pt} Between 5 and 9 years    & 4 \\
		\bf \hspace{5pt} Between 1 and 4 years    & 1 \\
		\bf \hspace{5pt} Less than 1 year         & 1 \\
		\midrule
		\bf Education &  16 \\
		\bf \hspace{5pt} Doctoral degree      & 2 \\
		\bf \hspace{5pt} Master degree        & 1 \\
		\bf \hspace{5pt} Bachelor degree      & 10 \\
		\bf \hspace{5pt} Diploma              & 2 \\
		\bf \hspace{5pt} Vocational certificate     & 1 \\
		\midrule
		\bf Perceived importance of search scenarios & 64 \\
		\bf \hspace{5pt} Essential & 20 (31.2\%) \\
		\bf \hspace{5pt} Very important & 25 (39.1\%) \\
		\bf \hspace{5pt} Moderately important & 16 (25.0\%) \\
		\bf \hspace{5pt} Somewhat important & 1 (1.6\%) \\
		\bf \hspace{5pt} Not important & 2 (3.1\%) \\
		\midrule
		\bf Urgency of obtaining an answer & 64 \\
		\bf \hspace{5pt} Extremely urgent (day) & 7 (10.9\%) \\
		\bf \hspace{5pt} Very urgent (days) & 22 (34.4\%) \\
		\bf \hspace{5pt} Urgent (week) & 20 (31.2\%) \\
		\bf \hspace{5pt} Somewhat urgent (weeks) & 9 (14.1\%) \\
		\bf \hspace{5pt} Not urgent & 6 (10.9\%) \\
		\bottomrule
	\end{tabular}
	}
\end{table}

Table~\ref{tbl:survey_stats} shows some statistics from the survey. The total number of users who participated in the survey was 16, divided among grain crop specialists (9) and agronomists (7). The majority had at least 10 years of experience and a bachelor degree. They tended to search for information that had significant bearing on farm or crop success with 70\% of the search scenarios either essential or extremely important. They also tended to be patient with their search with 65\% accepting to obtain an answer within days, up to a week. This suggests that answering agricultural users' information needs might be a slow search scenario, where you trade-off speed in favour of a high quality search experience~\cite{teevan2013slow}.

\begin{figure}
	\centering
	\includegraphics[width=\columnwidth]{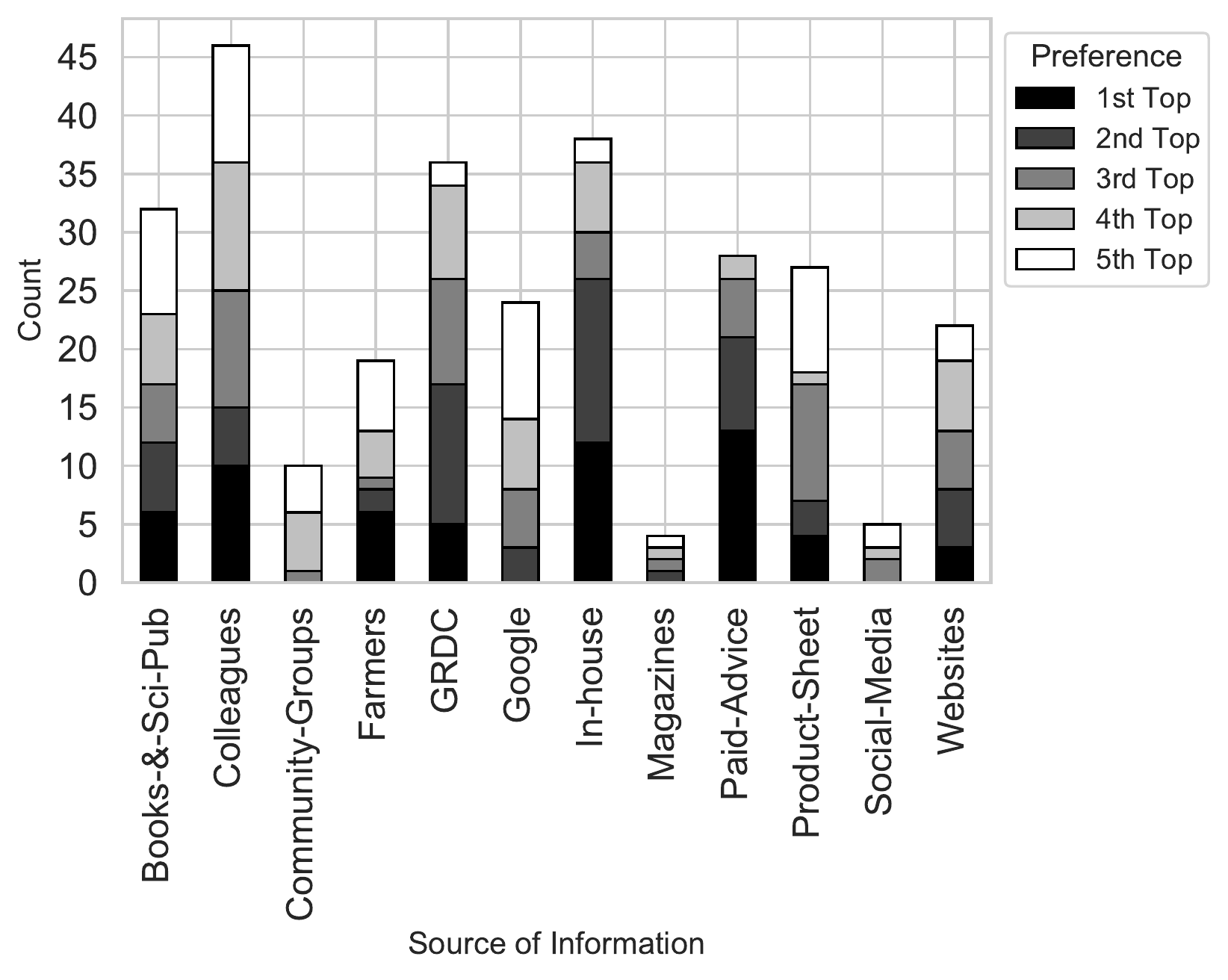}
	\caption{Preference for sources of information (frequency).}
	\label{fig:source_of_info}
\end{figure}

Figure~\ref{fig:source_of_info} depicts the preference for sources of information. We asked the users to select the top 5 out of 12 sources of information they would go to in order to help them find an answer to their questions. Agricultural experts tend to seek information from a wide range of sources with different levels of preference. They tend to trust more in-house reports (generated within the same organisation), colleagues, scientific publications, and paid advice. This is contrary to previous research that suggests Twitter is a popular platform for farmers to keep informed of the latest trends~\cite{mills2019use,van2011iphone}. This finding in specific, along with expert feedback from agronomist researchers in academia, government department and a leading research organisation, has influenced the selection of information sources for the documents collection.

\begin{figure}
	\centering
	\includegraphics[width=\columnwidth]{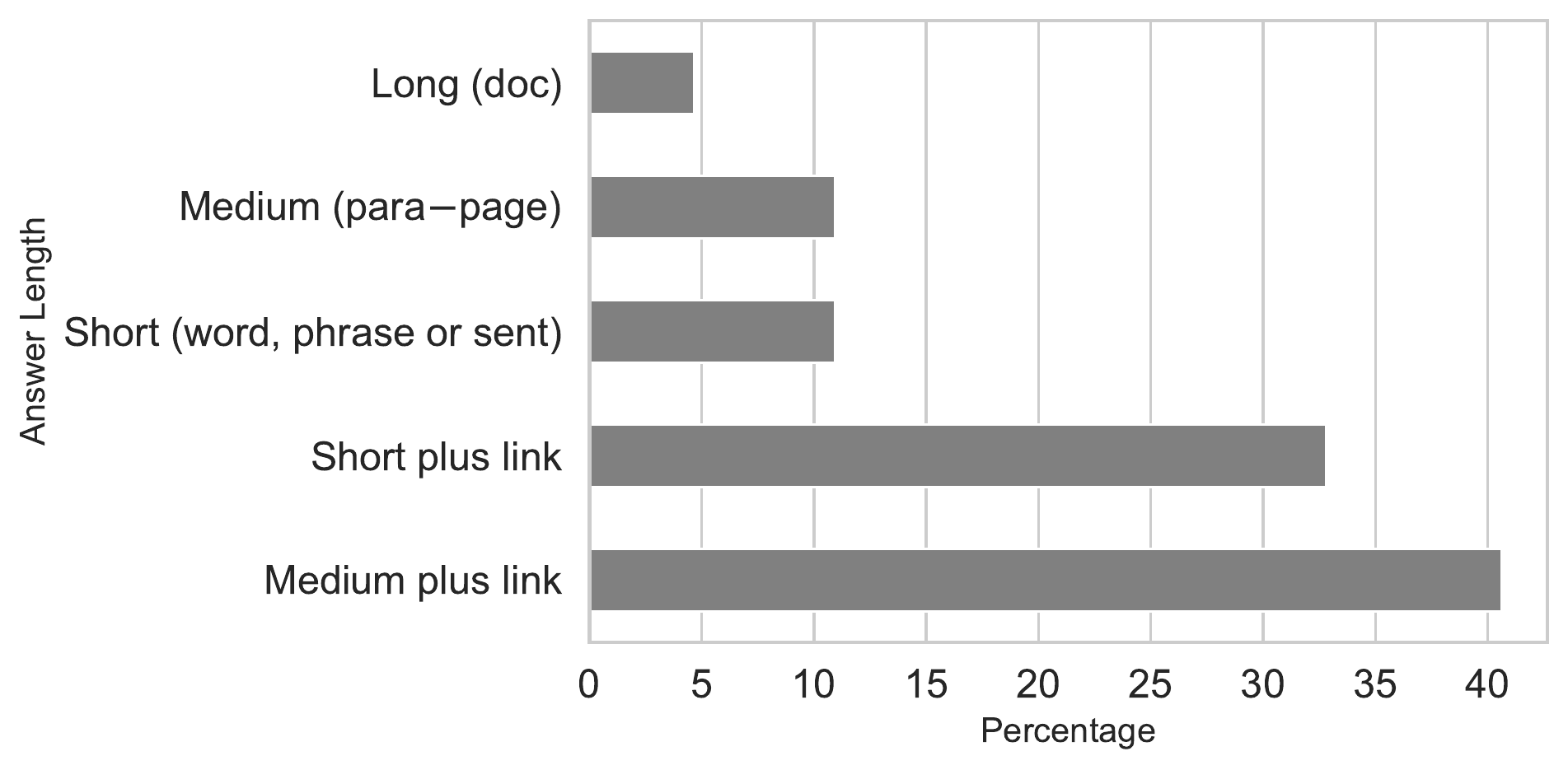}
	\caption{Answer length (percentage)}
	\label{fig:answer_length}
\end{figure}

\begin{figure}
	\centering
	\includegraphics[width=\columnwidth]{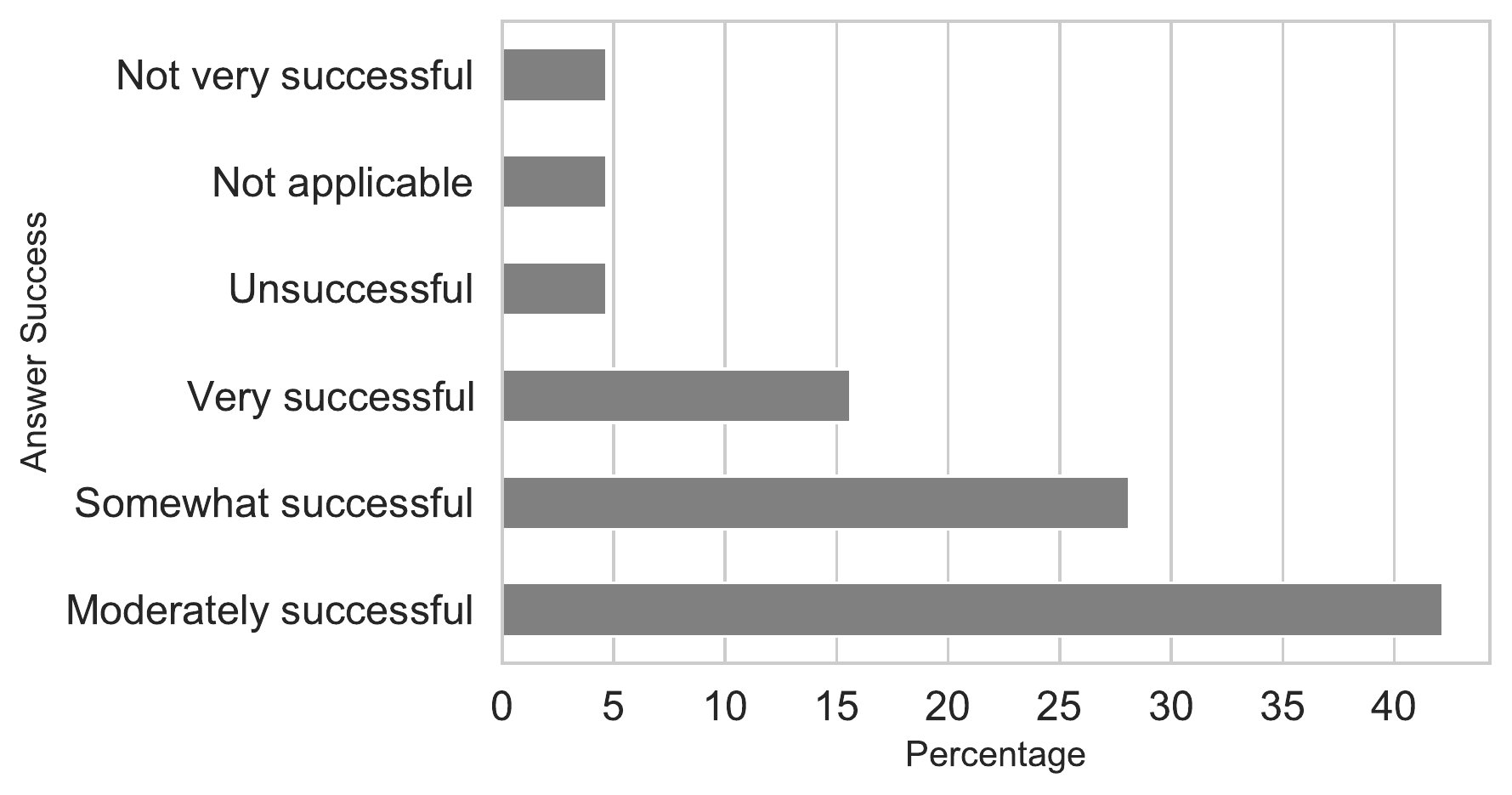}
	\caption{Answer success (percentage).}
	\label{fig:answer_success}
\end{figure}

Figure~\ref{fig:answer_length} shows the preference for the amount of information users would like to receive in the answer for each search scenario (Q7 in Table~\ref{tbl:sample_search_scenario}). Agricultural experts tend to prefer either short (single word, phrase or sentence) or medium (between a paragraph and a page) length answers with links to the evidence for further reading if required. Figure~\ref{fig:answer_success} demonstrates how successfully was an answer provided for each search scenario (Q10 in Table~\ref{tbl:sample_search_scenario}). Agriculture experts were very successful for only 15\% of the time. More than 65\% of the answers provided were somewhat (a partial answer was provided) or moderately (a good answer was provided, although they would have preferred more information) successful. These findings influenced the decision to formulate the task as passage retrieval where the passage meets the need of the moderate answer length and retrieval allows providing the source document of the answer.

%\add{What happened to the search scenarios and why did we drop them?}

%\add{Answer details -- they highlight the importance of medium length answers (lfqa, length of a passage) plus links for further evidence (retrieval augmented), hence passage retrieval}
%
%\add{Answer success -- only 15\% are successful. If the expert users (agronomists and specialists) are not able to find the answer, it's highly likely that the farmers will not be able to.}


\section{Test Collection} \label{sec:collection}
A key resource underpinning information retrieval research in a new area is an appropriate test collection for developing and evaluating new methods~\cite{voorhees2005trec}.
Creating the test collection involved three main steps: 1) creation of a set of topics (i.e., questions/queries); and 2) forming a pool of passages for judging; and 3) human relevance assessment. 

\subsection{Forming the Question/Query Topics}

%It was important that the topics were derived from the real information needs of users. Face-to-face interviews were not possible during the Covid pandemic and the online survey generated a small number of search scenarios. Therefore, we devised two alternative sources of real questions. 

%First, we used questions submitted by growers as part of an online webinar to a leading agriculture research organisation. These questions were focused on gaining recommendations for practice, and/or gaining the understanding of key factors to inform growers' management decisions. A sample question was ``How much stored soil water do I need to make best use of rainfall in my grain crops this year ?''.

%Second, to supplement the webinar questions 

We had two human assessors --- both agricultural scientists --- create questions via a process called known-item retrieval. The process for creating known-item topics was as follows:
\begin{enumerate}
	\item A random document was selected from the collection and shown to the assessor. If the document was not suitable for generating a reasonable question, the assessor select another random document.
	\item On reading the document, the assessor was asked ``What question does this document help answer''. The question they provided became the natural language question for a new topic.
	\item They were asked to provide 3 or more (unlimited) ad-hoc, keyword search queries that correspond to the question.
	\item They were asked to author an answer, in their own words, to that question.
	\item They were asked to select and paste the relevant portion of the document that helped answer the question.
	\item They were presented with a list of other passages from that document and asked to assess these as either relevant, marginal or non-relevant.
\end{enumerate}
A sample topic created using the above method is shown in Figure~\ref{fig:sample_topic}. A total of 210 topics were created from 165 documents (multiple, different topics could sometimes be derived from a single document).

Topics were divided into training and test sets. The 50 topics with the most relevance assessments formed the test set and the remaining 160 topics formed the training set. (Other splits can be done as desired; ours was purely done for our later experiments.)

\begin{figure*}
\small
  \ra{1.2}
  \begin{tabularx}{1\textwidth}{l|X}
    \toprule
    \bf Question: & \ttfamily What type of herbicides are effective against sowthistle? \\[7pt]
	\bf Keyword queries: & \ttfamily sowthistle herbicide mixing \\
    & \ttfamily Balance A Group D Group K sowthistle \\
    & \ttfamily broadleaf active herbicides mixing \\[7pt]
    \bf Assessor Authored Answer & \ttfamily The addition of Balance to either Group D or Group K herbicides can provide good control of sowthistle. The addition of Flame, Group D, Balance or Group K to broadleaf active herbicides (Group C and Valour) are also effective. \\[12pt]
    
    \bf Relevant Passages: & 20171829-40 (Relevant): \\
    & \ttfamily In the trials reported here the addition of Balance to either Group D or Group K has provided good control of sowthistle, when these same products applied alone are not providing acceptable control. \dots \\ % The addition of Flame, Group D, Balance or Group K to the broadleaf active herbicides (Group Cs and Valor) often improved their efficacy on sowthistle, but more importantly will improve control of the grass weeds common in the northern grains region. The long term residual control previously observed from Flame in grasses (greater than 150 days), was not observed for sowthistle in these trials. \\[7pt]
    & 20171829-39 (Marginal): \\
    & \ttfamily The trials reported here demonstrated that these products can perform quite poorly on the broadleaf weed sowthistle, when applied alone. The most effective products for sowthistle in these trials were Valor \dots \\ % and the two Group C products, but are limited by their shorter residual active life and, Group C in particular, can be quite weak in controlling grasses (Valor was not included in the 2016 trials). In 'Queensland Grains Research 2016', there was a consistent benefit in grass weed control from mixing two products with different modes of action. \\
    & \dots \\
    \bottomrule 
  \end{tabularx}
  \caption{Sample topic from the test collection. Each topic contains a question, a number of ad-hoc queries, an answer authored by assessors and a list of relevant passages graded with ``Relevant'', ``Marginally Relevant'' and ``Not Relevant''.}
  \label{fig:sample_topic}
\end{figure*}

\subsection{Documents and Passages}

Two sources of agricultural information were obtained as part of the collection: 4,003 agricultural reports from GRDC and the State Departments of Agriculture in Australia; and 82,843 scientific journal and conference articles from 33 agricultural journals.\footnote{Agricultural scientists and authors Y.Dang and D.Lawrence compiled a list of relevant journals.} These selected reports and journal articles were considered relevant to the grains industry and focused on crop agronomy and soils. The targeted subject matter related to the growth and management of grains crops including cereals (e.g., wheat, barley, and sorghum), legumes (e.g., chickpea, soybean, mungbean), and oilseeds (e.g., canola), and the management of the soils on which these crops are grown. Topics covered included recommendations and research relevant to the management of individual crops through varietals selection, sowing times, planting rates and row spacing etc; whole farming system performance, crop sequencing and fallow management practices; fertiliser management; and the identification and management of pest and diseases that affected the  grains industry. Both these sources came in the form of PDF documents.

The PDF reports we collated are open sourced\footnote{\url{https://doi.org/10.48610/fa4684b}}. The journal articles come from subscription journals so cannot be redistributed; however, we provide crawler scripts that can be used to download the full text using an institutional or paid subscription to these journals.

Once full-text PDFs were obtained, they were converted from PDF to JSON using Apache Tika. (Code for this is provided in the collection repository so that the processed collection can be fully reproduced, along with the pre-processed JSON files for the reports.) From here, the documents were further split into passages of three sentences (the Spacy sentencizer was used to derive sentence boundaries and code is provided for this.) From the 86,846 documents, 9,441,693 passages were produced.

\subsection{Pooling and Relevance Assessment}
\label{sec:pooling}

Using the 210 topics we set out to form a high quality pool for relevance assessment. We considered two state-of-the-art neural ranking systems, which we also used then for experimentation: monoBERT and TILDEv2. These models are described in Section~\ref{sec:retrieval_methods}.

Runs for all 210 topics were produced for each of the two systems above. These runs were fused using reciprocal rank fusion to produce the final pool for human assessment. 

\begin{figure*}
  \includegraphics[width=1.0\textwidth]{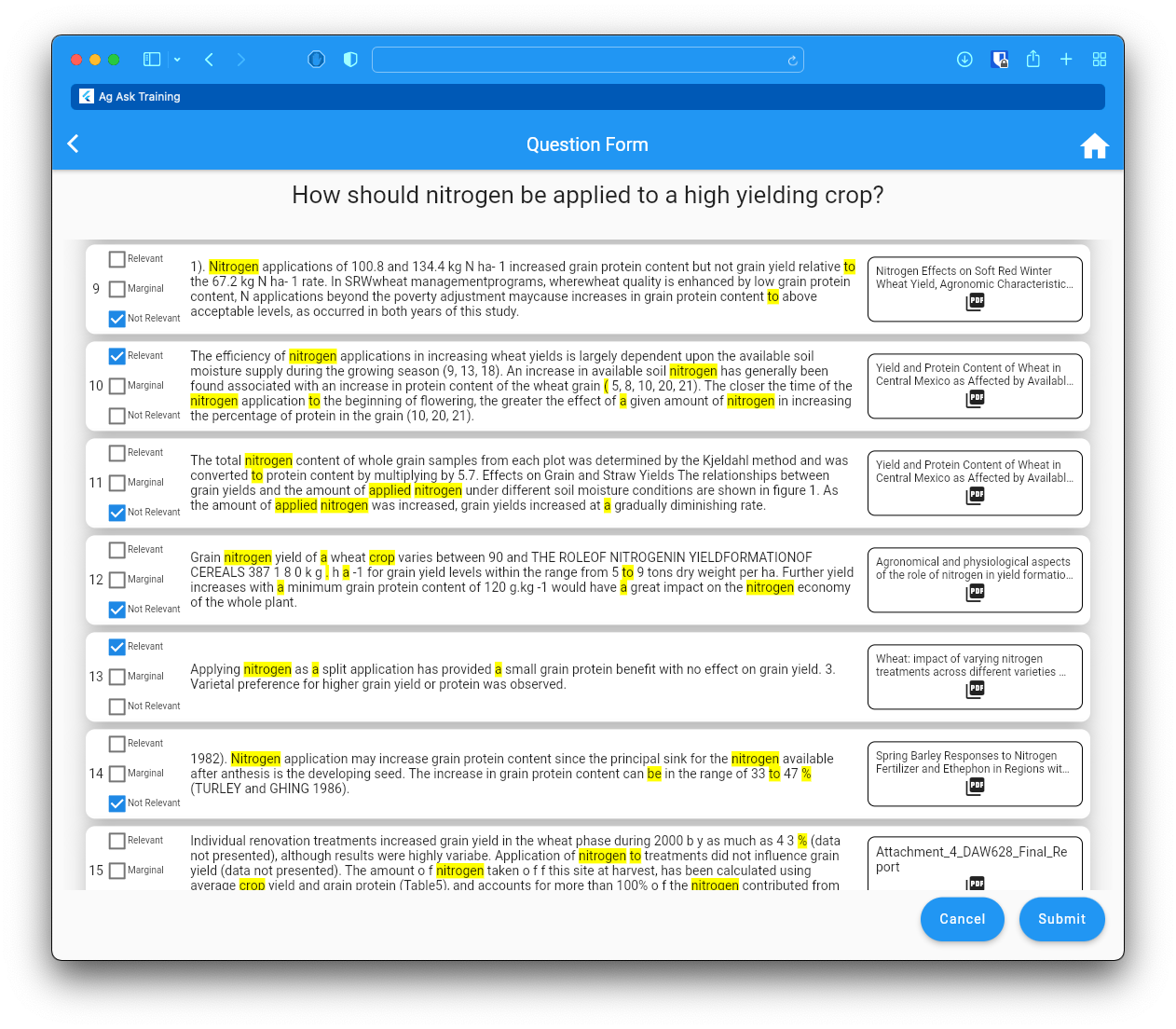}
  \vspace{-20pt}
  \caption{Relevance assessment of passages using the Agotator tool. The yellow highlighting indicates shared terms from the question and was provided to aid with relevance assessment.}
  \label{fig:Agotator-QuestionForm}
\end{figure*}

Relevance assessment was conducted by authors D.Lawrence and Y.Dang, both agricultural scientists. We developed a custom software tool called Agotator to support accurate and rapid relevance assessment.\footnote{We plan to open source Agotator in a future work.} A screenshot is shown in Figure~\ref{fig:Agotator-QuestionForm}.

As seen from the screenshot, users were presented with the topic question, a list of passages for judging, along with a link to the PDF source document from which the passage was extracted. Grades of relevance were: relevant, marginal and non-relevant. The criterion for relevance given to assessors was: ``does the passage help to answer the question'', where `relevant' meant that the passage contained the answer, `marginal' meant  the passage contained some part but not the whole answer, and `non-relevant` meant the passage contained no useful information.

For the topics from the test set, assessors judged in order until rank 20; if no relevant passage was found in the top 20 then they continued down the ranking until a relevant passage was found or rank 100 was reached.

For the topics from the training set, assessors judged the top 10 passages, regardless of relevance. Topics obtained with this known-item retrieval process will have at least 1 relevant passage.

\subsection{Characteristics of the Test Collection}

Table~\ref{tbl:collection_stats} provides statistics for different parts of the test collection. Topics in the collection were multi-faceted, containing a natural language question, a number of keyword queries, a human authored answer, and relevance passages; a sample topic is shown in Figure~\ref{fig:sample_topic}.

\begin{table}[t]
  \ra{1.1}
  \centering
  \caption{Statistics of the test collection we compiled.}
  \label{tbl:collection_stats}
  \begin{tabular}{lr}
    \toprule
    \bf Topics & 210 \\
    \bf \hspace{5pt} Train & 160 \\
    \bf \hspace{5pt} Test & 50 \\
    \midrule
    \bf Judged Passages & 3948 \\
    \bf \hspace{5pt} Non-relevant & 1244 (32\%) \\
    \bf \hspace{5pt} Marginal relevant & 852 (22\%) \\
    \bf \hspace{5pt} Relevant & 1852 (48\%) \\
    \midrule
    \bf Documents & 86,846 \\
    \bf \hspace{5pt} Reports & 4,003 \\
    \bf \hspace{5pt} Journal articles & 82,843 \\
    \bf Passages & 9,441,693 \\
    \bottomrule
  \end{tabular}
\end{table}

\begin{figure}[t]
  \includegraphics[width=1.0\columnwidth]{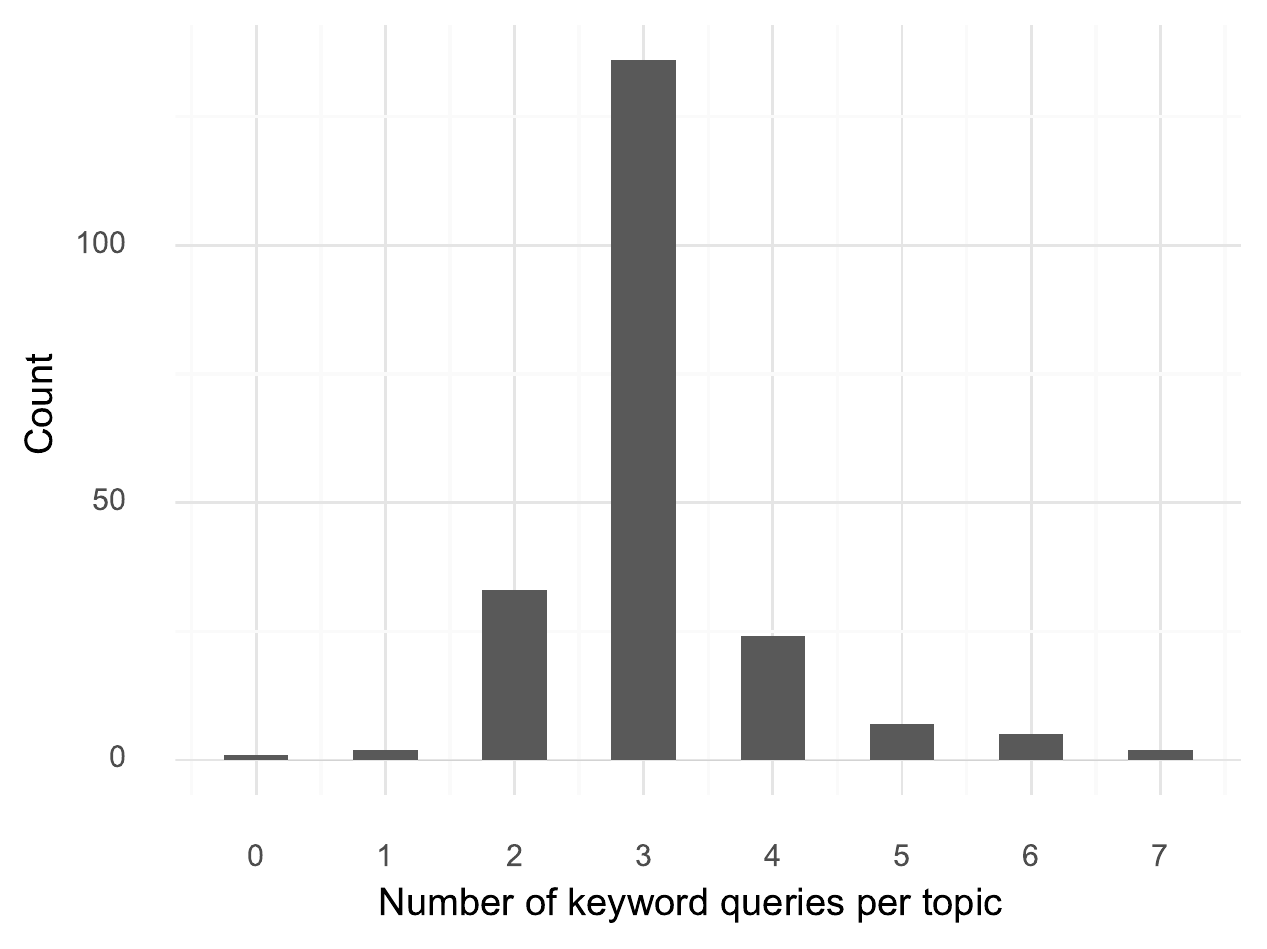}
  \caption{Histogram showing the number of keyword queries for each topic. Mean=$3$, SD=$0.92$.}
  \label{fig:queries_per_topic}
\end{figure}

\begin{figure}[t]
  \includegraphics[width=.9\columnwidth]{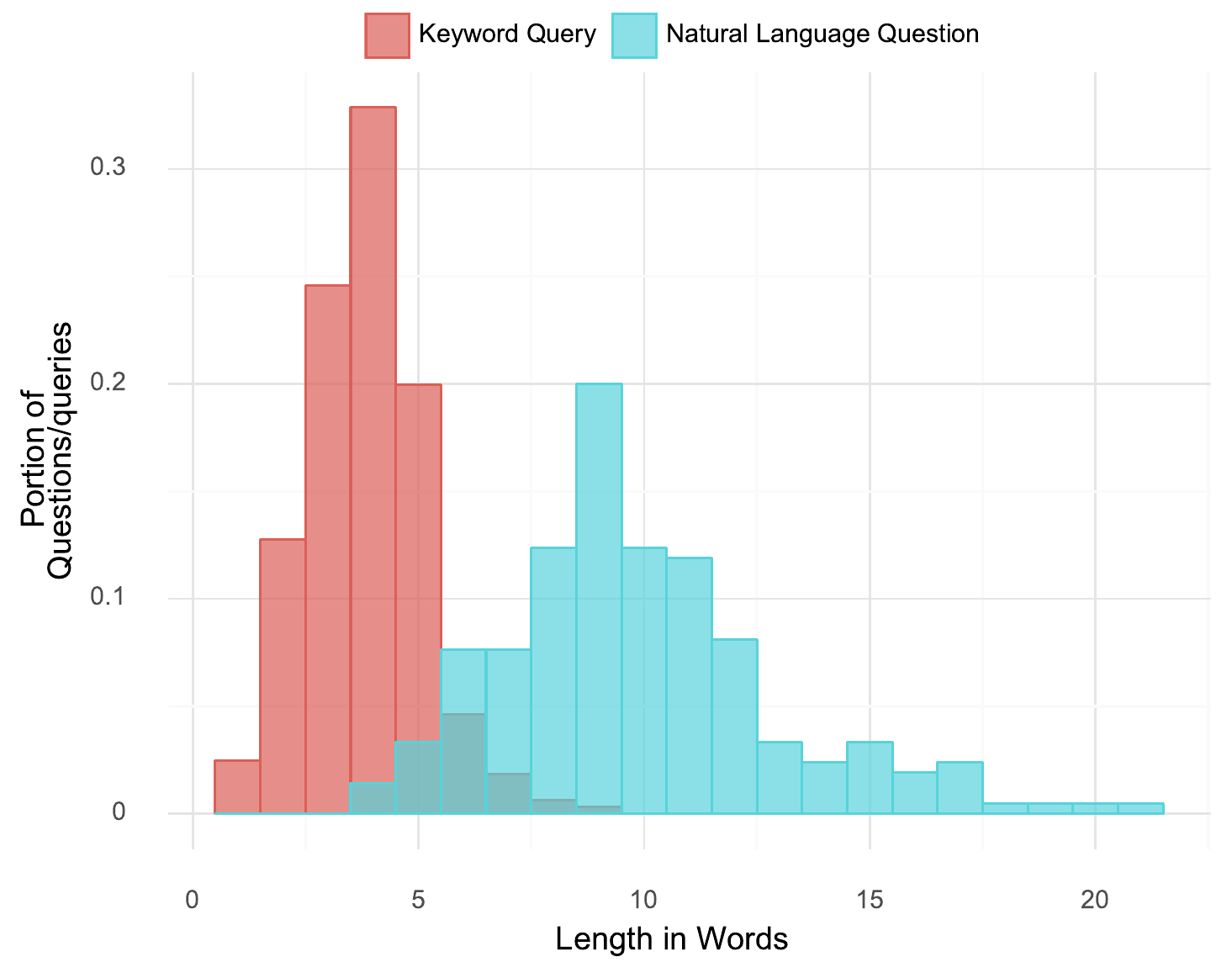}
  \caption{Query length in number of words for natural language questions and keyword queries. Mean length for questions = $9.8$ words and keywords = $3.8$.}
  \label{fig:query_length}
\end{figure}

As seen, the test collection supports query variations by having multiple keyword queries for each topic. Figure~\ref{fig:queries_per_topic} shows a histogram of the number of keyword queries for each topic. Most topics contain three queries (mean=$3$, SD=$0.92$), as per the instructions to assessors to provide at least three. Query length in number of words is shown in Figure~\ref{fig:query_length}. As to be expected, natural language questions were both longer and more varied in length. Most keyword queries were between 3 and 4 words long.

\begin{figure*}[t]
	\centering
	\includegraphics[width=0.7\textwidth]{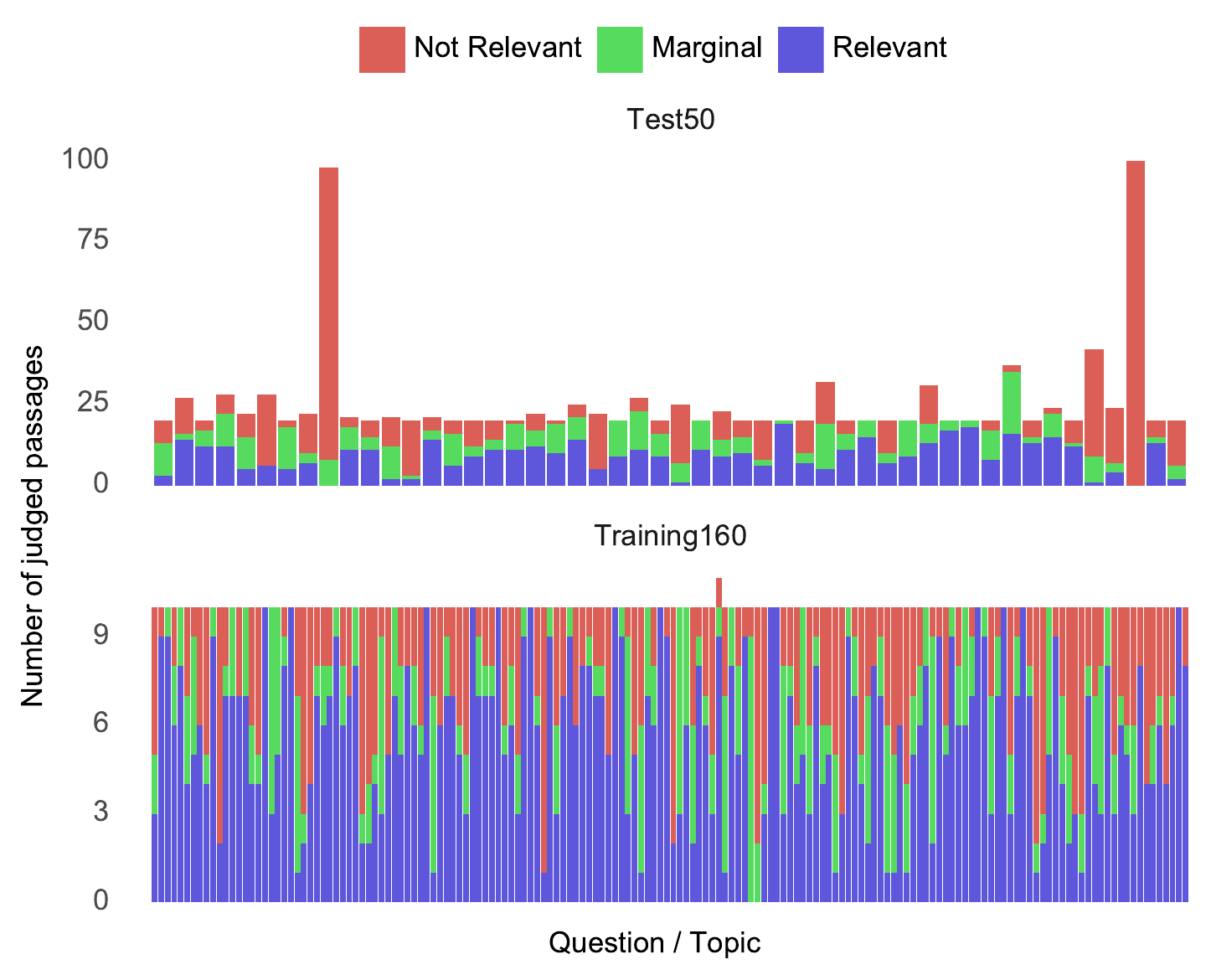}
	\caption{Breakdown of relevant and non-relevant passages for each question in the training and test sets.}
	\label{fig:qrels}
\end{figure*}

Figure~\ref{fig:qrels} shows the breakdown of grades of relevance for the topics in training and test sets. Recall that for the test set, assessors judged to rank 20, stopping there if at least one relevant passage was found, otherwise continuing down the ranking until a relevant passage was found or rank 100 reached. As seen from the plot, no relevant passages were found for one topic. We opted to keep this topic in the collection because it was from the known-item retrieval set, which means there was at least one relevant passage, but that had not been been retrieved by any of our models in the pool.

\section{Passage Retrieval}
\label{sec:experiments}

%In this section, we conduct initial experiments with \ag to reveal both the characteristics of the collection and to demonstrate its utility. These are, of course, not exhaustive of what \ag can be used for and merely serve as a primer.

Two main experiments were conducted: 1) understanding the effectiveness of a selection of retrieval models on this collection; 2) understanding how query variations impact effectiveness.

\subsection{Retrieval Methods}
\label{sec:retrieval_methods}

We implemented the following retrieval methods: %and evaluated them on the \ag  test collection:

%[style=unboxed, leftmargin=0.5cm]
\begin{itemize}
	\item \textbf{BM25:} Vanilla BM25 baseline to understand how a simple term-based retrieval performs.
	\item \textbf{BM25-RM3:} A BM25 baseline with pseudo relevance feedback using RM3.
%	\item[monoBERT Reranker:] BM25 followed by a monoBERT reranker pre-trained on MSMARCO and fine-tuned on the 160 training topics. (The same system used for pooling.)
%	\item[TILDEv2 Tuned:] The same computationally efficient neural document expansion model used for pooling. 
%	\item[TILDEv2:] TILDEv2 without fine-tuning on the target domain, providing an estimate of the benefit of performing fine-tuning.
%%	\item[ANCE:] A dense retriever that selects more realistic negative training instances from an Approximate Nearest Neighbor (ANN) index of the corpus~\cite{Xiong:2020ww}. We used ANCE model pre-trained on the MSMARCO dataset. It helps to evaluate the generalisability of \ag to other retrieval models that are not considered for pooling.
%%	\item \add{Add ANCE fine-tuned on questions}
%	\item \add{Add monoBERT and TILDEv2 fine-tuned on queries} %  and ANCE 
	
		\item \textbf{monoBERT}: a cross-encoder neural method involving a first stage BM25 initial retrieval of 1000 documents, followed by a fine-tuned monoBERT reranker~\cite{nogueira2019multi}. We used a monoBERT model pre-trained on the MSMARCO dataset and then fine-tuned on the 160 training topics.
	
	\item \textbf{TILDEv2}, is a neural reranker that utilises document expansion at indexing time to avoid the need for neural encoding of documents at query time~\cite{Zhuang:2021vt}. It involved a first stage BM25 retrieval of 1000 documents, followed by a fine-tuned TILDEv2 reranker. TILDEv2 was added as a computationally efficient --- yet still effective --- model that might be deployed in a live search system. This model was also fine-tuned on the 160 training topics.
\end{itemize}

To make use of the multi-faceted topics provided in the collection, we ran the above models using both the natural language questions and keyword query versions of the topic. This aimed to uncover some insights into how query variation impact effectiveness.

\subsection{Results}

\begin{figure*}
  \includegraphics[width=1.0\textwidth]{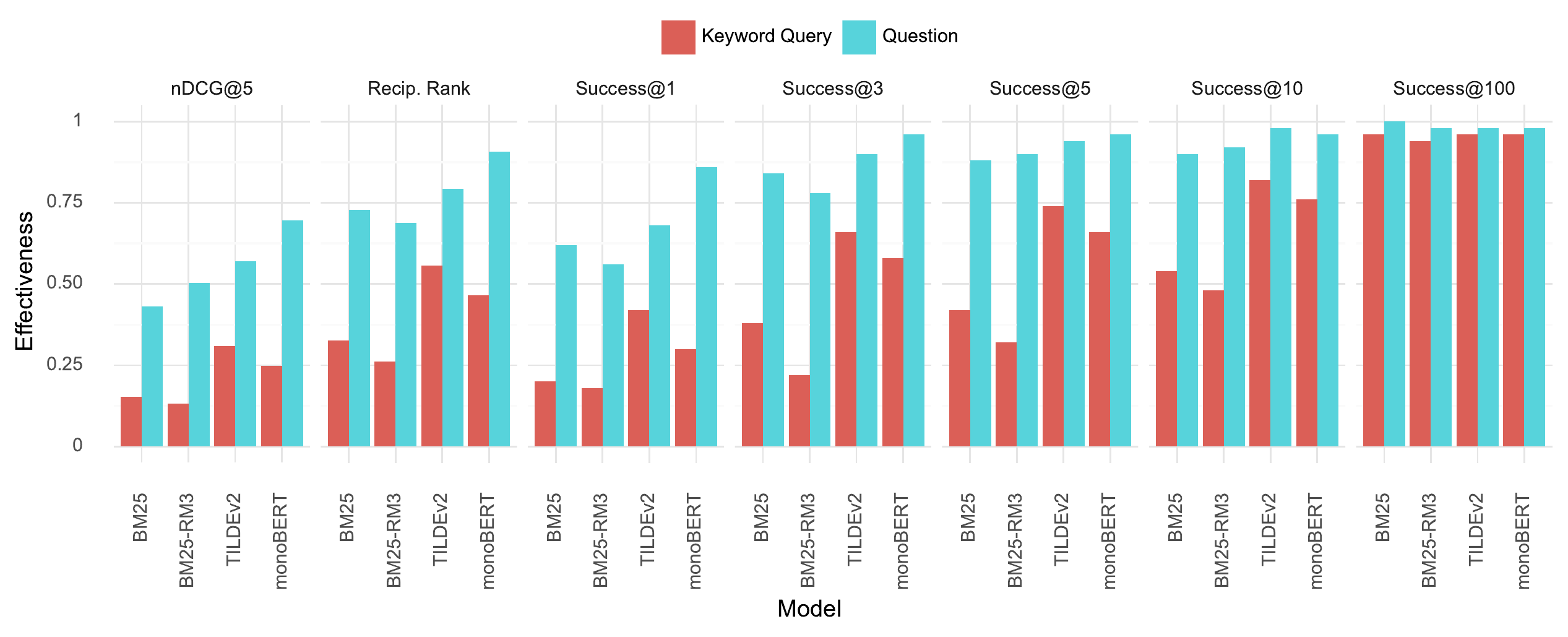}
  \caption{Retrieval effectiveness of different models. Both natural language questions and  keyword query topic types were evaluated.}
  \label{fig:ijdl_model_effectiveness}
\end{figure*}

The effectiveness of the above models are shown in Figure~\ref{fig:ijdl_model_effectiveness}.

\subsubsection{Term-based vs neural model effectiveness}

There was a large difference in effectiveness between the term-based BM25 model and the neural rankers on this collection: monoBERT and TILDEv2 models were far more effective than BM25 (t-test, $p<0.01$ for nDCG@5). However, it's worth noting that for measures like Success@100, BM25 was highly effective (no statistically significant difference between BM25 and neural models). This meant that BM25 \emph{retrieved} the relevant passages, but was not effective at \emph{ranking} them (low effectiveness for measures that consider top ranked results; e.g., NDCG@5). This tells us that using a BM25 for initial retrieval was reasonable, if it was followed by a high-precision reranker.

\subsubsection{Natural language vs. keyword queries}

Using natural language questions was more effective than keyword queries in most cases. (This is somewhat contrary to pervious research that has shown verbose queries are less effective~\cite{Bendersky:2008wm}.) The neural models, in particular, were suited to questions rather than queries. The benefit of using questions is seen in early precision and not recall; i.e., improvements were seen in measures such as NDCG@5 and reciprocal rank that measure early precision rather than success@100 that measure recall. 

The popular technique of pseudo relevance feedback on top of BM25 (i.e. RM3) actually reduced effectiveness for keyword queries. However, pseudo relevance feedback was effective when applied to natural language questions.

\section{End to End Integrated Solution: AgAsk}

\begin{figure}
	\centering
	\includegraphics[width=1\linewidth]{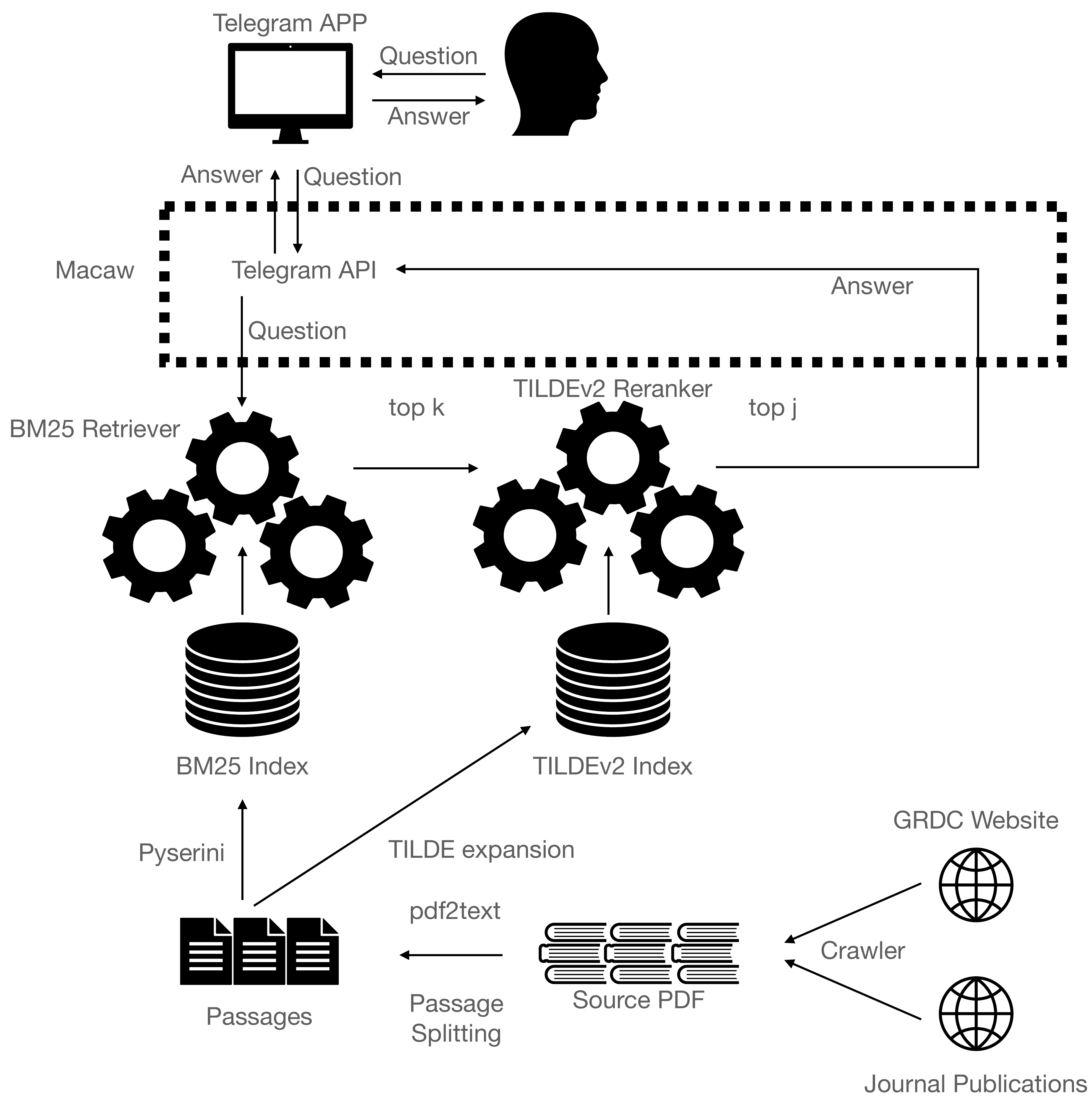}
	\caption{Overall architecture of AgAsk.}	\label{fig:architecture}
\end{figure}

In this section, we describe our single entry-point system for agricultural users to help them search for information, dubbed \ak. AgAsk can be deployed as a conversational agent, or a traditional search engine.
Figure~\ref{fig:architecture} provides the overall architecture of AgAsk in its deployment as a conversational agent. A grower uses Telegram to ask his question to the `AgAsk' bot. Overall conversation management is handled by Macaw~\cite{zamani2020macaw}, an open-source framework for building conversational search systems. Macaw passes the query to our custom retrieval pipeline, comprising of a first stage BM25 retriever and the neural TILDEv2 re-ranker~\cite{Zhuang:2021vt}. Retrieved passages are then sent to Macaw, which is responsible for serving it back to the grower via Telegram.

%to the BART answer generation model which converts the passages into a single coherent answer. The answer is then fed back to Macaw, which is responsible for serving it back to the grower via Telegram.

\subsection{Client and User Interface}

%\begin{figure}
%	\centering
%	\includegraphics[width=0.85\columnwidth]{figures/agask}
%	\caption{\vspace{-10pt} A screenshot of the AgAsk Telegram bot.}
%	\label{fig:agaskbot}
%\end{figure}

\begin{figure}[t]
	\centering
	\includegraphics[width=1\columnwidth]{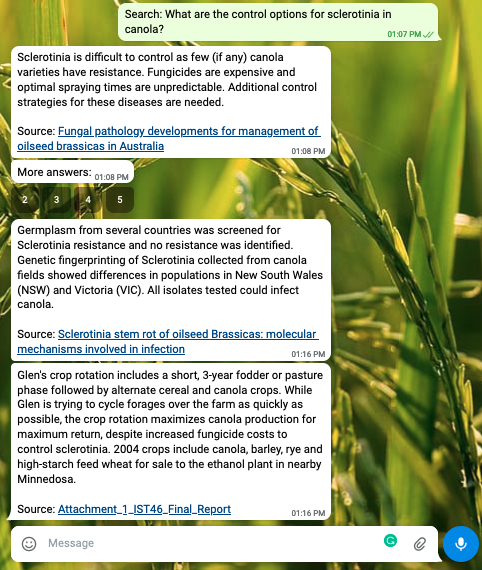}
	\caption{A screenshot showing a AgAsk in use. Top is the user's question, along with best match answer passage. Buttons under "More answers:" allow the user to see the next four ranked passages, two of which are displayed. Each document from which the passage was extracted is shown as a hyperlink next to "Source:".}
	\label{fig:agaskbot}
\end{figure}

AgAsk is accessible to agricultural users via Telegram; an example screenshot is shown in Figure~\ref{fig:agaskbot}. %\footnote{Note, the question clarifications in the screenshot are obtained in a Wizard-of-Oz setting. The screenshot shows the output of the passage retrieval component, which is fully automated.}. 
Telegram was chosen because it provides a simple API and Telegram clients are available for every major platform and device. 
The grower can pose a natural language question and AgAsk will respond with a generated answer. 

A demonstration video of AgAsk is available at \url{https://ielab.io/projects/agask.html}. The retrieval of passages is done by the AgAsk bot. The clarifying questions are currently manually inserted to demonstrate what a fully interactive system might look like. We are in the early stages of deploying in production such a mixed-initiative conversational system.

We also log all user interactions including clicks, likes and emojis; this provides a source of relevance feedback information that may be used in future feedback mechanisms or online learning to rank.

\subsection{Conversation Management with Macaw}

AgAsk employs the Macaw conversational information seeking framework~\cite{zamani2020macaw}, as it provides a convenient way of building an entire pipeline from scratch.
The Macaw framework consists of several modules, including intent identification, co-reference resolution, query generation, retrieval model, and result generation. Currently, we have disabled the intent identification, co-reference resolution, query generation, file IO, and standard command line IO modules. We have instead instantiated our own retrieval and result generation modules, as detailed above, while we are in the process of deploying in production relevant modules for intent identification, relevance feedback, and question clarification.
%The input of Macaw can also be in different formats, file IO, standard command line IO, and Telegram. 

%Then, we discuss the traditional trade-off between efficiency and effectiveness in real time systems and the criteria considered for the selection of the currently deployed retrieval model.

\subsection{Choices in Retrieval Model}

The monoBERT reranker was the best performing model (see from Section~\ref{sec:experiments}). If you consider a live question-answering system that might provide three possible answers to a user's question (e.g., in a conversational or mobile setting) then success@3 would be the measure to consider. In this setting monoBERT provided a success@3 of 0.96: 48/50 topics had a relevant passage in the top 3 results. We posit this would make for a highly effective real system if the results generalise beyond the test topics in our collection.

\begin{figure}[t]
	\centering
	\includegraphics[width=\columnwidth]{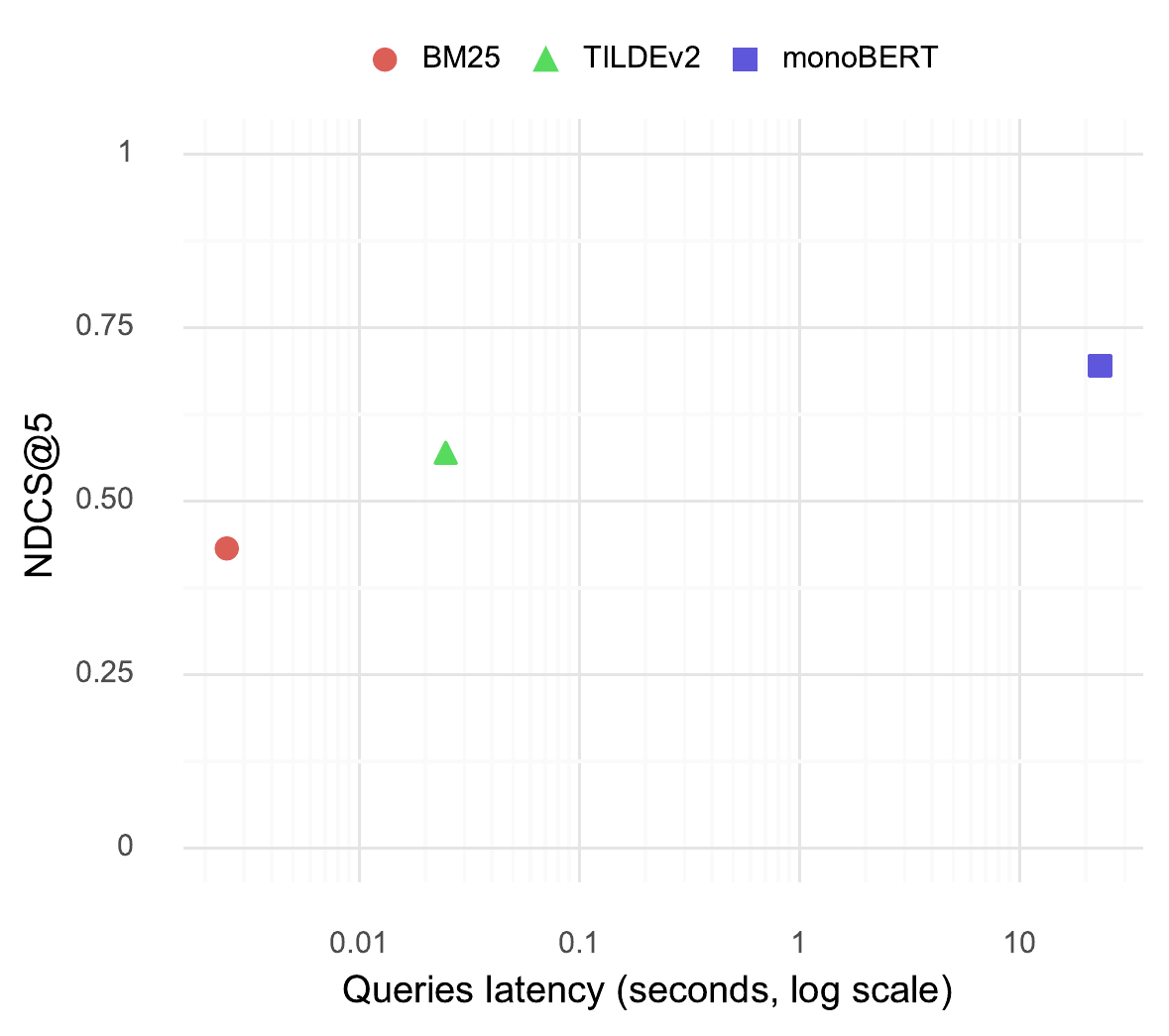}
	\caption{The effectiveness-efficiency tradeoff for different retrieval models for AgAsk. While monoBERT is the most effective, it cannot serve queries to a user in a timely manner. TILDEv2 offers a far more efficient model with only a small reduction in effectiveness; hence TILDEv2 was chosen as the underlying model for AgAsk.}
	\label{fig:ijdl_query_latency}
\end{figure}

While monoBERT was highly effective, it was computationally expensive. Query latency would make it prohibitive for real users in an online passage retrieval setting; or specialist GPU and parallel hardware might be required. TILDEv2, while less effective, was far more efficient, and could be deployed in production on commodity CPU-based hardware (although document expansion and indexing were best done using a GPU).

Figure~\ref{fig:ijdl_query_latency} depicts the effectiveness-efficiency tradeoff for different retrieval models for AgAsk. It suggests that achieving more effectiveness requires more query latency. This is particularly evident when comparing monoBERT to either BM25 or TILDEv2. monoBERT achieves a higher NDCG@5 with a considerable trade-off in latency. On the other hand, TILDEv2 strikes a great balance between effectiveness and query latency. Hence, we employ TILDEv2 in \ak. A further advantage of using TILDEv2 is that it does not need a dedicated GPU-based server to be used in production, as monoBERT does instead, as TILDEv2 runs entirely on CPU for its inference stage.

\section{Future Work}

\subsection*{Further research using the test collection}

The test collection detailed in Section~\ref{sec:collection} is a standalone resource that can be used, independent of the AgAsk system, in the development and evaluation of search systems for the agricultural sector.

\textbf{Passages vs documents}: The collection contains both full documents and sub-document passages. This allows other researchers to investigate differences in effectiveness between passage and document retrieval~\cite{kaszkiel1997passage,liu2002passage,craswell2021overview}.

\textbf{Query variation}: Topics in the collection are multi-faceted: they contain a natural language question and multiple keyword queries. Query variations have a large impact on retrieval effectiveness~\cite{moffat2015pooled} and the study of query variation is an active research area~\cite{bailey2016uqv100}. The test collection provides a query variation resource. The fact that it contains both natural language questions and keyword queries means that these two different representations can be analysed by others.

\textbf{Answer generation}: For each topic, assessors authored an answer to the topic question in their own words. (The sample topic from Figure~\ref{fig:sample_topic} shows this.) Note that these answers may differ in vocabulary or substance from relevant passages from the document. They represent the assessors expression of what the document contains. Using this, the collection could be used to develop and evaluate answer generation methods that, for example, take a set of retrieved passages to derive a natural language answer to the question~\cite{hsu2021answer}. The answers could also be analysed to understand how similar or different the language used in answers is to that of relevant passages.

\textbf{Scientific document extraction}: The reports and journal articles used in the collection were processed using a basic PDF extraction method that divided the document into three sentence passages. This method did not account for the structure of the document --- paragraphs, figures, tables, sections, etc. Using the collection, one could investigate many different information extraction methods for scientific articles~\cite{bast2017benchmark}, and the impact that these have on retrieval, answer generation or any other downstream tasks.

\textbf{Domain specific / expert search}: Previous research has de\-monstrated the value of and the need for domain-specific test collections to evaluate the effectiveness of open-domain information retrieval models~\cite{salampasis2013integrating,tait2014introduction,verberne2019first,thakur2021beir}. However, there is no search datasets for agriculture. Our test collection represents a domain-specific, expert scenario. Models that work in the open domain may not translate to this expert domain. The test collection provides a resource to test this, and possibly develop new models suited to this domain.

\subsubsection*{The case for contextualisation}

AgAsk matches a query to a passage without taking into account characteristics of the user --- there is no personalisation or contextualisation. Through our analysis of the information needs of users in agriculture, we identified that certain characteristics of a user have a strong bearing on their information need and impact on what they would consider relevant answers to their questions. We detail some below. We further note this need for contextualisation and adaptation to the different practices of the individual users within the same professional domain is a common characteristic across other professional search tasks~\cite{russell2018information}.

\textbf{Weather \& Climate}: Information should be tailored to recent weather, forecasted and longer term climatic predictions (e.g., if the farmer is located in a drought predicted area then recommendations for drought resistant crops would be important).

\textbf{Location}: The grower's region strongly informs their information need. The growing conditions, access to markets, infrastructure (e.g., rail or irrigation networks), historical crop yields and many other factors can be inferred from location. Thus, growers would like information that is location-aware.

\textbf{Markets}: Contextualisation to the specific market that the grower operates in, including price, trends and changing customer demands/preferences.

\textbf{Literacy/Interpretably}: Evidence-based agriculture involves making decisions based on scientific evidence and sources. While growers may recognise the value of this, they do not necessarily want to delve into detailed scientific information, or have the expertise to do so. Instead, they would like outcomes of the scientific literature to be provided to them in an understandable, concise and digestible form. Furthermore, grower's expertise varies considerably --- some may have detailed technical expertise in certain areas and thus would like to see associated technical details; others may have no technical expertise in the area and require a lay overview. Information should be tailored to different grower's literacy and expertise.

If the above information about the user was available to a search system such as AgAsk, then the retrieval model could take this into account when ranking passages. In Telegram, this information could be recorded as part of a user's profile. How to use this information in one of the retrieval models (e.g., TILDEv2), is an open and interesting area of future work.

\section{Conclusion}

This paper presents AgAsk --- a search system designed to answer farmers questions where information is extracted from scientific documents. While scientific documents on agriculture contain a plethora of useful information, they are not accessible or easily searchable by farmers with specific information needs. AgAsk attempts to overcome this by building a search system specifically for this problem.

Understanding the information needs of farmers is critical in designing a good search system to support them. We conduct a thorough analysis of information needs through a survey of users who were given real search scenarios to perform. This reveals the type of information they look for (e.g., crop protection, product recommendations) as well what source of information they use (books, Google, product sheets), what form they would like their answers as (e.g., a short answer with link to longer document). Learnings from this project informed the requirements for a search system and the basis of forming a test collection to evaluate such a system.

We form a test collection comprising 210 real questions, a collection of 86,846 scientific documents (split into 9,441,693 passages). Two agricultural experts did manual relevance assessment indicating which passages were relevant to each question. This provides ground truth for both training machine learning retrieval models and for empirical evaluation. The collection contains different query types (natural language vs keyword), as well as human generated answers; thus providing a resource for further research on query variations and automated answer generation.
The test collection is made public to foster further research into search in the agricultural domain.

Using the test collection we train and evaluate a number of passage retrieval models, including two state-of-the-art neural rankers --- TILDEv2 and monoBERT. An empirical evaluation of all methods shows that neural rankers can be highly effective at finding relevant passages to a farmer's question. 

%We tackle some key practical consideration of deploying search system by

How to deploy the above models in a usable system is often non-trivial. We describe a deployment architecture that makes use of the Telegram messaging platform for the front-end client and Macaw conversational search platform for the back-end server. This provides a flexible and scalable architecture. An analysis of the efficiency-effectiveness tradeoff of different retrieval models highlights how neural rankers such as monoBERT are not practical for deployment in live systems and thus alternative, non GPU models such as TILDEv2 are preferred.

Finally, we highlight how the agricultural domain offers an interesting test bed for further research, with a key focus on better personalisation/contextualisation (e.g., location or weather aware rankers). It is our aim to both foster more research in this area and to translate research into real-world systems deployed in the field.

Data, code and the test collection are available at: \url{https://github.com/ielab/agvaluate}.
\section*{Declarations}

Funding for this project was provided by the Grain Research Development Corporation under project\# UOQ2003-009RTX.
 
Ethics approval related to the survey we conducted was granted by The University of Queensland under application \#2020000826.

The authors have no competing interests to declare that are relevant to the content of this article.

\bibliographystyle{plain}
\bibliography{ijdl2022-AgAsk_end_to_end}% common bib file


\end{document}